\documentclass{article}

\usepackage[square,comma,numbers]{natbib}

\usepackage{tabularx}
\usepackage[utf8]{inputenc} 
\usepackage[T1]{fontenc}    
\usepackage{hyperref}       
\usepackage{url}            
\usepackage{booktabs}       
\usepackage{amsfonts}       
\usepackage{nicefrac}       
\usepackage{microtype}      
\usepackage{xcolor}         
\usepackage{amsmath} 
\usepackage{graphicx}
\usepackage{multirow}
\title{Reconstructing 3D Flow from 2D Data with Diffusion Transformer}

\author{%
  Fan Lei\\
    Institute of Artificial Intelligence\\
    Xiamen University\\
}

\begin{document}

\maketitle

\begin{abstract}
Fluid flow is a widely applied physical problem, crucial in various fields. Due to the highly nonlinear and chaotic nature of fluids, analyzing fluid-related problems is exceptionally challenging. Computational fluid dynamics (CFD) is the best tool for this analysis but involves significant computational resources, especially for 3D simulations, which are slow and resource-intensive. In experimental fluid dynamics, PIV  cost increases with dimensionality. Reconstructing 3D flow fields from 2D PIV data could reduce costs and expand application scenarios. Here, We propose a Diffusion Transformer-based method for reconstructing 3D flow fields from 2D flow data. By embedding the positional information of 2D planes into the model, we enable the reconstruction of 3D flow fields from any combination of 2D slices, enhancing flexibility. We replace global attention with window and plane attention to reduce computational costs associated with higher dimensions without compromising performance. Our experiments demonstrate that our model can efficiently and accurately reconstruct 3D flow fields from 2D data, producing realistic results. 
\end{abstract}

\section{Introduction}

Fluid flow is perhaps the most widely applied physical problem to date, from the design of aerospace vehicle wings \cite{reguly2015acceleration}, to optimizing the output rate of chemical reactors, to analyzing pollutant dispersion \cite{jimenez2013near} and wastewater treatment in environmental issues. Due to the highly nonlinear and chaotic nature of fluids, analyzing fluid-related problems becomes exceptionally challenging. Analytically solving the Navier-Stokes equations, which determine the essence of fluid behavior, remains an impossible task \cite{pope2001turbulent}.

Computational fluid dynamics (CFD) is the best tool for analyzing fluid flow. It uses numerical methods and computer simulations to solve fluid problems. A typical CFD simulation  process involves generating a mesh, i.e., dividing a geometric model into small discrete elements, setting boundary and initial conditions, and solving the Navier-Stokes equations to obtain flow field data.

Although CFD can simulate fluid flow quite accurately in most cases, it involves a significant investment of computational resources, especially for simulations in 3D space, and very slow. Common methods to reduce computational resource consumption include performing calculations only on half (or even a quarter) of the geometric model, assuming the model is highly symmetrical, or performing calculations in 2D space, using information from the 2D space to describe fluid flow characteristics in 3D space. These approaches involve strong assumptions that partial information of the flow field can capture the overall fluid flow characteristics and may require extensive specialized knowledge and manual effort. However, these assumptions may not always hold. Therefore, we often need to perform CFD simulations in 3D space to fully capture the fluid flow characteristics in the entire 3D space.

In the field of experimental fluid dynamics, particle-image velocimetry (PIV) \cite{adrian1984scattering,scarano2012tomographic,schanz2016shake} is a key method for obtaining flow field information. PIV uses high-speed cameras to capture particle images during laser pulses, recording particle positions in the flow field. Algorithms compare the particle distribution in different frames to find the distance particles move within the time interval, thereby calculating the velocity vector at each point in the flow field and generating velocity field data. The cost of PIV technology is also affected by dimensionality. For example, 3D3C PIV requires more high-speed cameras than 2D3C PIV to capture velocity field data in 3D space. Therefore, if 3D flow fields can be reconstructed from 2D PIV data, it will greatly reduce costs and expand the application scenarios of PIV technology.

The large amount of flow data obtained from numerical simulations and experiments directs our attention to data-driven methods. In the field of fluid dynamics, machine learning, especially deep learning methods \cite{lecun2015deep}, has made significant progress \cite{brunton2020machine,kutz2017deep,vinuesa2022enhancing,duraisamy2019turbulence,vinuesa2023transformative}, from improving measurement techniques based on machine learning \cite{guastoni2021convolutional,guemes2021coarse} to accelerating numerical solution methods \cite{tompson2017accelerating,obiols2020cfdnet}, surrogate models \cite{nakamura2021convolutional,yousif2022reduced}, flow field prediction and reconstruction \cite{yousif2023deep,lienenzero,drygala2023generalization}, and flow control \cite{fan2020reinforcement,han2020active}. Machine learning methods have shown strong application potential, with deep learning models accurately modeling highly nonlinear and chaotic fluids, making them an ideal choice for reconstructing 3D flow fields. Although deep learning models usually require a long training process, once trained, their inference speed can be several orders of magnitude faster than traditional numerical methods.

On the other hand, the vigorous development of generative models \cite{sohl2015deep,ho2020denoising,song2020score,dhariwal2021diffusion,rombach2022high} and the research progress in 3D reconstruction tasks \cite{gao2022get3d,mescheder2019occupancy,mildenhall2021nerf,jain2022zero,kerbl20233d} make it possible to reconstruct the 3D flows from some partial observations (such as 2D flow field). In this paper, we demonstrate that the state-of-the-art generative model has strong generative capabilities, it trained with flow data to acquire sufficient prior knowledge about the flow field, can recover the entire 3D flow field information from partial flow field information, thereby obtaining the flow characteristics of the 3D flow field. 
Specifically, we trained a Transformer-based Diffusion model \cite{vaswani2017attention,ho2020denoising,peebles2023scalable}, integrating plane position embeddings to capture the spatial information of 2D planes while utilizing window and plane attention mechanisms to handle the computational complexity of three-dimensional flows. This design enables the reconstruction of 3D flow fields from 2D flow data, with the plane position embeddings allowing the model to recover 3D flow information from arbitrary combinations of 2D planes.

In summary, the contributions of this paper are:
\begin{itemize}
    \item We propose a method for reconstructing the 3D flow field from partial flow data, e.g., 2D flow observations, leveraging the strong prior knowledge of a trained diffusion model.

   \item We replace vanilla global attention with the simpler yet highly effective window attention and plane attention to alleviate the increased computational cost brought by higher dimensions. Our results demonstrate that these new attention mechanisms can significantly reduce computational costs with minimal impact on model performance.
       
    \item By embedding the position information of 2D planes into the model, we enable the recovery of 3D flow fields through any 2D plane combination, effectively utilizing training data and enhancing the flexibility of 3D flow field reconstruction.
    
    \item We demonstrate through experiments that our model can successfully reconstruct 3D flow fields from 2D flow field data, capturing the true distribution of the flow field and generating realistic samples.
\end{itemize}

\section{Related Work}

\textbf{Flow Prediction and Reconstruction.}\quad Flow prediction focuses on forecasting flow behavior at future time steps based on domain geometry, boundary, and initial conditions. Although numerical solvers offer high accuracy, they are computationally expensive, with resource demands escalating rapidly in higher dimensions, driving the search for more efficient alternatives. Hybrid methods leverage data-driven models to replace costly components of numerical solvers, while purely machine learning approaches, such as neural networks, directly predict flow fields. Neural operators \cite{anandkumar2020neural,tranfactorized,kovachki2023neural,li2020fourier,li2022transformer,hao2023gnot} exemplify this by learning functional mappings for flow prediction. Other works applying deep learning to flow field prediction include \cite{kim2019deep,lienenzero,drygala2023generalization}. Additionally, incorporating physical constraints into models \cite{raissi2019physics,wandellearning} has proven to enhance prediction accuracy. Flow field reconstruction, in contrast, seeks to recover the full 3D flow field from partial observations, leveraging the model’s prior knowledge of the underlying flow dynamics. \citet{matsuo2021supervised} and \citet{yousif2023deep} are the most relevant to our work. \citet{matsuo2021supervised} utilized CNNs to reconstruct 3D flow fields from multiple parallel 2D sections, while \citet{yousif2023deep} employed a GAN-based approach to recover 3D flow fields from two orthogonal 2D flow fields. Unlike their approaches, we employ a powerful generative model, the diffusion model, combined with a Transformer architecture. 
Additionally, unlike their methods, which recover 3D flow fields from fixed-position 2D planes, we introduce a novel plane position embedding, allowing the model to reconstruct 3D flow fields from different combinations of 2D planes.

\textbf{3D Reconstruction.}\quad Recent advancements in computer vision have significantly enhanced 3D reconstruction techniques \cite{gao2022get3d,mescheder2019occupancy,mildenhall2021nerf,liu2023zero,shi2023zero123++,kerbl20233d,muller2024multidiff,niemeyer2024radsplat,zhang20233dshape2vecset}, with promising applications for 3D flow field reconstruction. Neural Radiance Fields (NeRFs) \cite{mildenhall2021nerf,jain2022zero,deng2023nerdi} generate 3D scenes by training neural networks on multiple 2D images to produce color and density outputs. Generative models further advance this field by directly creating 3D representations from 2D images \cite{liu2024one} or synthesizing novel views \cite{liu2023zero,shi2023zero123++,muller2024multidiff} to support 3D scene reconstruction \cite{voleti2024sv3dnovelmultiviewsynthesis}. Additionally, significant progress has been made in Gaussian splatting \cite{niemeyer2024radsplat,kerbl20233d,yu2024mip} and implicit representations \cite{tang2021sa,tretschk2020patchnets,zhang20223dilg,zhang20233dshape2vecset}. Among these methods, the most relevant to our 3D flow field reconstruction is the generative model-based approach. However, instead of generating novel object views from images, we reconstruct the corresponding 3D flow field from provided 2D flow field slices. In this work, we represent the 3D flow field as voxels and reconstruct it from 2D slices. Common 3D representations include point clouds \cite{vahdat2022lion,nichol2022point,melas2023pc2,cao2024motion2vecsets}, meshes \cite{liu2023meshdiffusion,liu2021deepmetahandles,gao2022get3d}, and voxels \cite{gadelha20173d,sanghi2022clip,mo2023dit}.  Of these, voxel-based techniques are particularly relevant to this approach. However, other 3D representations may also provide viable solutions for flow field reconstruction.

\textbf{Diffusion Models.}\quad Generative models, particularly diffusion models \cite{sohl2015deep,ho2020denoising,song2020score}, have advanced rapidly, emerging as state-of-the-art in generation quality and diversity. They now outperform other generative frameworks, such as variational autoencoders (VAEs) \cite{kingma2013auto}, autoregressive models (ARMs) \cite{chen2020generative}, flow-based models \cite{dinh2014nice}, and generative adversarial networks (GANs) \cite{goodfellow2020generative}. Diffusion models have shown great potential across fields, with recent applications to 3D data \cite{cao2024motion2vecsets,mo2023dit,voleti2024sv3dnovelmultiviewsynthesis,nichol2022point,shi2023zero123++,muller2024multidiff}. While U-Nets \cite{ronneberger2015u} have traditionally served as the backbone for diffusion models, Transformer-based architectures are gaining traction \cite{vaswani2017attention}. DiT \cite{peebles2023scalable} employs a pure Transformer backbone, demonstrating impressive scalability. Extensions like DiT-3D \cite{mo2023dit}, U-ViT \cite{bao2023all}, and UniDiffuser \cite{bao2023one} further explore Transformers in diffusion models. Given their success, we selected Transformers for reconstructing 3D flows. To manage the computational complexity of 3D spatial data, we replaced global attention with plane and window attention. These linear-complexity attention mechanisms introduce inductive biases by leveraging the locality of flow fields, enhancing efficiency without sacrificing expressive power. 

\section{Method}

\subsection{Preliminaries}\label{Prelim}
In this subsection, we first introduce the problem setting of reconstructing 3D flow fields from 2D flow fields. Then, we briefly provide some background knowledge on denoising diffusion probabilistic models (DDPMs). Finally, we describe two-stage approach and CFG, explaining our reasons for not using them.

\textbf{Problem Setup.}\quad For a given set of 2D slices \( P = \{ p_i \}_{i=1}^n \) within a 3D flow field, each 2D slice \( p_i \in \mathbb{R}^{d_1^i \times d_2^i \times c} \), where \( c \) represents flow field attributes such as pressure or velocity, our goal is to reconstruct the 3D flow field \( S \in \mathbb{R}^{d_x \times d_y \times d_z \times c} \). We aim to learn \( f_\theta \), a neural network parameterized by \( \theta \),
\begin{equation}\label{model}
    \hat{S}=f_{\theta } (P, E_P),
\end{equation}
as a function of the 2D plane set \( P \) and the plane position embedding \( E_P \) (section \ref{PPE}), and outputs the reconstructed 3D flow field \( \hat{S} \). We want \( \hat{S} \) to be as similar as possible to \( S \).

To create \( f_{\theta } \) for reconstructing the flow field with high quality, we need to address two key issues. First, reconstructing a 3D flow field from 2D slices is very challenging, as it is an ill-posed problem that involves inferring global information from local data and requires strong prior knowledge.
The second question is how do we address the potentially enormous computational costs associated with 3D data? For the first problem, we chose a generative model with sufficient generalization capabilities—a diffusion transformer. It learns the prior knowledge about flow fields from the training data, as introduced in section \ref{Diffusion}. For the second question, we addressed it by replacing the traditional attention mechanism with more computationally efficient window attention and plane attention mechanisms. Furthermore, to enhance the flexibility of flow field reconstruction,we propose the plane position embedding, discussed in section \ref{PPE}, which allows us to fully utilize the 3D data.

\textbf{Background on Diffusion Models.}\quad Diffusion Models define the forward process as a Markov chain, which incrementally injects noise into the original data. The reverse process is also a Markov chain, continuously removing noise from a certain prior distribution to recover the original data. The essence of Diffusion Models lies in learning this reverse process.

For the original 3D flow field data \( S_0 \), the forward process continuously adds noise, generating \( S_1, S_2, \ldots, S_T\). The typical form of the transition kernel is usually  \(q(S_t | S_{t-1}) = \mathcal{N}(S_t; \sqrt{1 - \beta_t} S_{t-1}, \beta_t \mathbf{I}),\)
 where \(\beta_t \in (0, 1)\) is a Gaussian noise value.
 
For the backward process, The reverse Markov chain is characterized by a transition kernel \( p_{\theta}(S_{t-1} | S_t) = \mathcal{N}(S_{t-1}; \mu_{\theta}(S_t, t), \Sigma_{\theta}(S_t, t))\), where \(\theta \) denotes model parameters. For training, we can form variational upper bound \cite{kingma2013auto} which reduces to \( \mathcal{L}(\theta)=-p(S_0 | S_1) + \sum_{t} D_{KL}(q(S_{t-1} | S_t, S_0)  ||  p_{\theta}(S_{t-1} | S_t))\). 
Given that both \( q(S_t | S_0) \) and \( p_{\theta}(S_{t-1} | S_t) \) are Gaussian, and considering that the backward process operates conditioned on the 2D flow field and its corresponding positional embedding, we can simplify the training objective to
\begin{equation}
    \mathcal{L}_{simple}(\theta) = \mathbb{E}_{t, S_0, \epsilon} \left \| \epsilon - \epsilon_{\theta}(S_t, c(t, P, E_P)) \|^2_2 \right .
\end{equation}
where \( c(t, P, E_P) \) is the condition of timestep, 2D flow field and plane position embedding. Once \( p_{\theta}(S_{t-1} | S_t) \) is trained, the model \(f_{\theta}\) can reconstruct the 3D flow field \( S_0 \) by applying iterative denoising conditioned on \(c(t, P, E_P)\) .

\textbf{Two-stage Approach and CFG.}\quad Due to the substantial computational resources required to train diffusion models in high-resolution pixel/voxel space, LDM \cite{rombach2022high} employs a two-stage approach to address this issue: (1) Train a high-quality autoencoder and use its encoder \(E\) to map the data \(S\) to a lower-dimensional latent space. (2) Freeze the autoencoder and train the diffusion model in the latent space to restore the latent representation of the raw data from Gaussian noise \(z_T\) to \(z_0 = E(x)\). After model training is completed, we can sample in the latent space and use the decoder to obtain the original data \(S = D(Z)\).

LDM can achieve better 3D flow field reconstruction with fewer computational resources. However, we did not use it for two reasons: (1) There is no off-the-shelf autoencoder trained on 3D flow field data. (2) Training a 3D autoencoder for 3D flow from scratch requires a large amount of high-quality flow field data and computational resources. Training a diffusion model directly in the voxel space is less costly. Therefore, we have set training the autoencoder as a task for future work.

Classifier-free guidance (CFG) is commonly used in conditional diffusion models. According to Bayes Rule, \( \log p(c|x) \propto \log p(x|c) - \log p(x) \), so \( \nabla_x \log p(c|x) \propto \nabla_x \log p(x|c) - \nabla_x \log p(x) \). Here, \( c \) represents conditional information, such as 2D flow information. Through reparameterization, \( \hat \epsilon_{\theta}(x_t, c) = \epsilon_\theta(x_t, \emptyset) + s \cdot \nabla_x \log p(x|c) \propto \epsilon_\theta(x_t, \emptyset) + s \cdot (\epsilon_\theta(x_t, c) - \epsilon_\theta(x_t, \emptyset)) \), where \( s > 1 \) is the guidance scale, \(\emptyset\) represents unconditional generation. By this, we can sample from \( p(x|c) \) to obtain \( x \) under the condition \( c \). However, we found in experiments that the reconstruction results of the model are highly sensitive to \( s \). Therefore, to achieve accurate 3D flow reconstruction and avoid hyperparameter tuning, we did not use CFG.

\subsection{Plane Position Embedding}\label{PPE}
To enable the model to reconstruct 3D flow fields from any combination of 2D flow fields, we input both the features of the two-dimensional flow fields and their position embedding into the model. We use a Cartesian coordinate system to describe the spatial relationships within the flow field. For convenience, we normalize the coordinates of each point in 3D flow field to be contained inside the XYZ unit cube \( [0,1]^3\). Consider the equation of planes \(Ax + By + Cz + D = 0, \) we normalize it by dividing by \(\mathbf{n}  = \sqrt{A^2 + B^2 + C^2}\), resulting in \(A'x + B'y + C'z + D' = 0.\)We do not consider planes where \(D' > 1\), thereby ignoring planes that intersect the cube \( [0,1]^3\) with a small area. Instead of directly inputting the four values \(A'\), \(B'\), \(C'\), and \(D'\) into the model, we feed them into the model via Fourier feature embeddings. This approach enriches the features and provides a continuous and smooth representation of plane equation.

Specifically, for the given dimension \( d \), and plane equation parameter \( A' \), the embedding can be represented as
\begin{equation}\label{PPE1}
    \text{PE}(A', 2i) = \sin \left( A'/10000^{(2i / d)-1} \right),
\end{equation}
\begin{equation}\label{PPE2}
    \text{PE}(A', 2i+1) = \cos \left( A'/10000^{(2i / d)-1} \right).
\end{equation}
where \(i\) is the dimension index.
Note that since \(A'\), \(B'\), \(C'\), and \(D'\) are less than 1, we multiply them by 10,000. Consequently, in Equation (\ref{PPE1}) and (\ref{PPE2}), the exponent of the numerator becomes \((2i / d) - 1\) instead of the usual \(2i / d\). We concatenate the embeddings of the four plane equation parameters to obtain the positional embeddings \(E_{p_i} = [\text{PE}(A'_{i}),\text{PE}(B'_{i}),\text{PE}(C'_{i}),\text{PE}(D'_{i})]\) for each plane \(p_i\). We then concatenate these to form the overall plane positional embedding \(E_P = [E_{p_1},\dots ,E_{p_n}]\), project it to the model dimension \(D\) using a multi-layer perceptron (MLP), add it to the timestep embedding \(t_{emb}\), and feed the result into the model.

\begin{figure*}[t]
    \centering 
    \includegraphics[width=12cm]{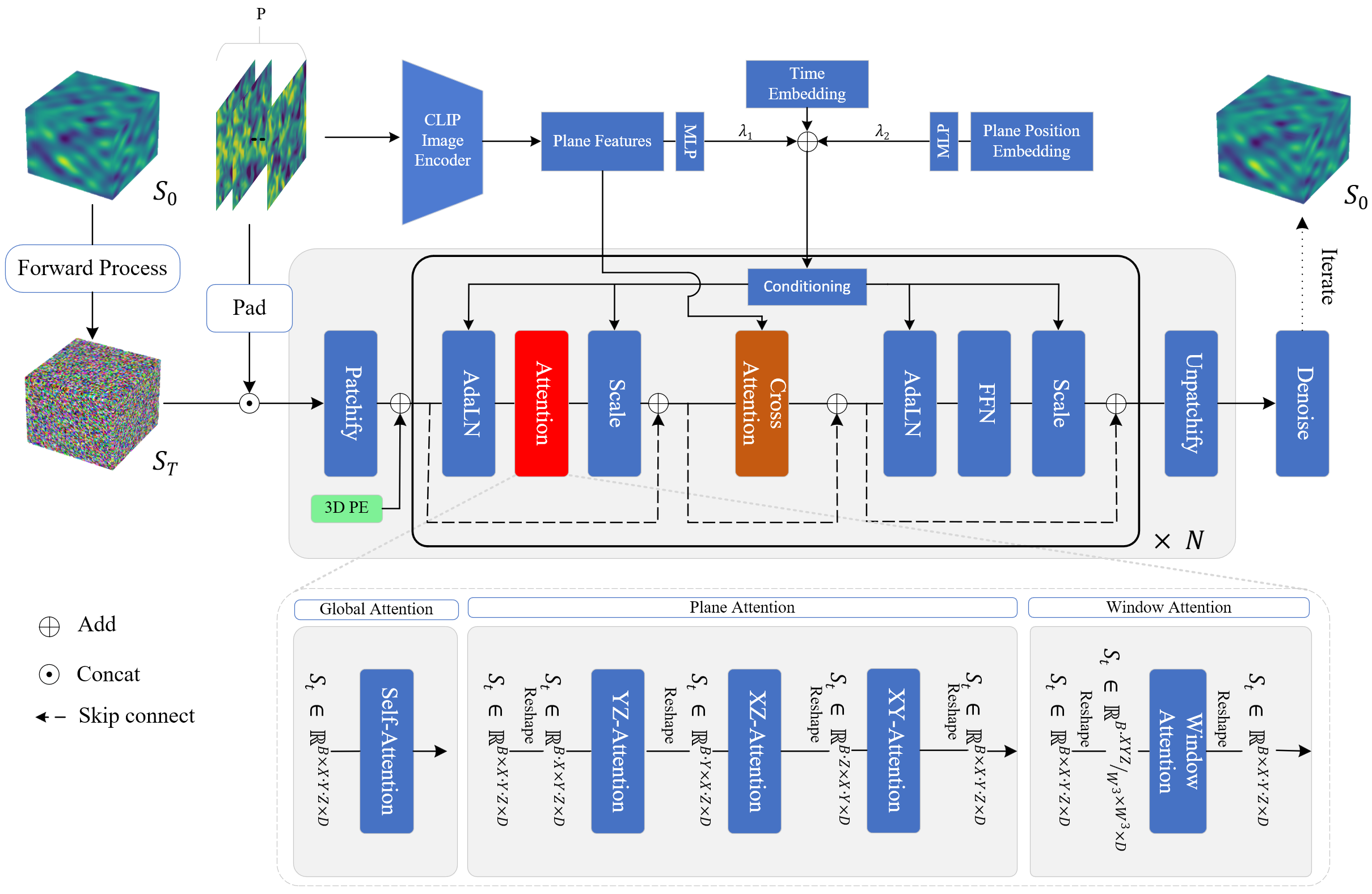}
    \caption{\textbf{Pipeline and architecture.} We use the Diffuse Transformer to reconstruct 3D flow from 2D flow. \textbf{Top}: The 2D flow is fed into the model through three paths: 1) it is padded and concatenated with the input; 2) features are extracted using CLIP, combined with the timestep embedding and plane position embedding, and used as conditioning; and 3) the final layer output of the CLIP image encoder is fed into the model via cross-attention. \textbf{Middle}: The structure of the Diffuse Transformer. The input flow field undergoes patchification, passes through \(N\) transformer layers, and is unpatchified to produce the predicted noise in the flow field. Global attention is replaced with more efficient window attention and plane attention within the transformer. \textbf{Bottom}: The structure of different attention mechanisms. Both window attention and plane attention are implemented through reshaping operations.}
    \label{fig:pipeline}
\end{figure*}

\subsection{Diffusion Transformer for 3D Flow Field Reconstruction}\label{Diffusion}
Inspired by recent works \cite{mo2023dit, peebles2023scalable, chenpixart}, we use a diffusion transformer to reconstruct 3D flow from 2D. The 3D flow field is converted into input tokens through a 3D patchify operation, and 3D positional embeddings are added before processing the tokens through \(N\) transformer layers, which utilize window and plane attention. The 2D flow features are extracted using CLIP which are input to the model through adaptive layer norm. Finally, a linear layer and an unpatchify operation output the predicted noise in the 3D flow. The entire pipeline and architecture are shown in Figure \ref{fig:pipeline}.

\textbf{3D Patchify.}\quad The typical Transformer architecture processes data of shape \(B\times L\times D\), where \(B\) denotes the batch size, \(L\) is the sequence length, and \(D\) is the dimension of the model. However, we treat the 3D flow field as voxels, \( S \in \mathbb{R}^{d_x \times d_y \times d_z \times c} \). Therefore, the first layer of our model is designed to patchify \(S\). For a patch size of \(p_x \times p_y \times p_z\), we use a 3D convolution layer with both kernel size and stride set to the patch size. This layer converts the input flow field voxels into a sequence of 3D patchified tokens \(T_{S} \in \mathbb{R}^D\) with sequence length \(L = X\cdot Y \cdot Z = \frac{d_x}{p_x}\cdot \frac{d_y}{p_y}\cdot \frac{d_z}{p_z}\). The patch size does not impact the number of parameters in the model but affects the size of \(L\), which is the attention window size in the transformer. Changes in patch size influence \(L\) and model's flops and, consequently, the quality of flow reconstruction \cite{peebles2023scalable}.
Next, we add 3D positional embeddings \cite{mo2023dit} to the input tokens based on the spatial position of the corresponding voxels. 

\begin{figure*}[tb]
    \centering 
    \includegraphics[width=12cm]{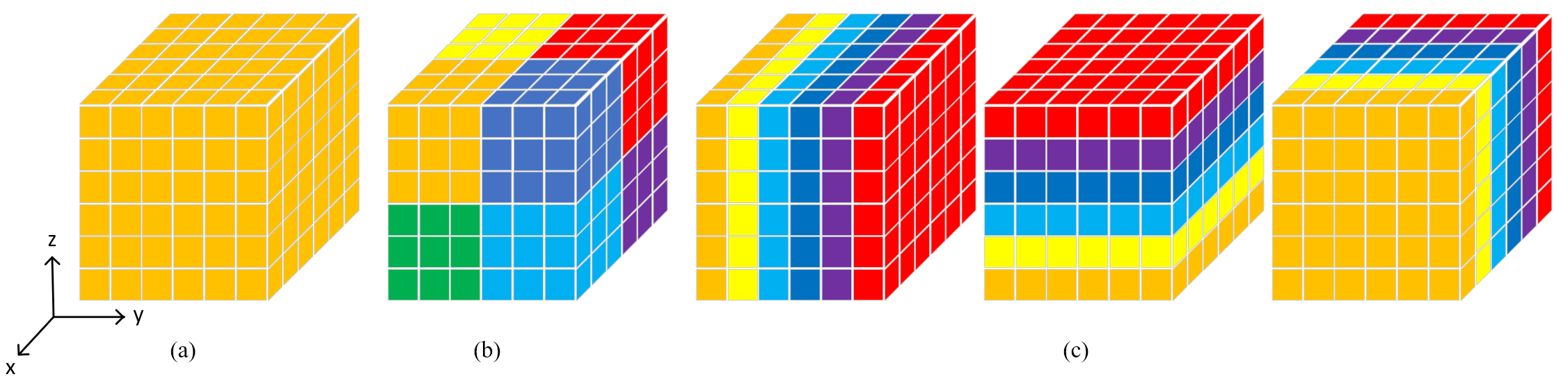}
    \caption{\textbf{Visual explanation of different attention.} Each grid of the cube represents an input token, and only tokens of the same color can attend to each other. (a) Global attention. (b) Window attention. (c) Plane attention: from left to right, \(yOz\) 
plane attention, \(xOz\) plane attention, and \(xOy\) plane attention.}
    \label{fig:attention}
\end{figure*}

\textbf{Transformer Layers.}\quad Each transformer layer contains a self-attention layer, a cross-attention layer ,and a feed-forward layer. 
The self-attention is defined as \( \text{attention}(x_i) = \text{softmax}\left(\frac{\mathbf{Q}\mathbf{K}^T}{\sqrt{D}}\right)\mathbf{V} \), where \( \mathbf{Q} = x_i\mathbf{W}^\mathbf{Q} \), \( \mathbf{K} = x_i\mathbf{W}^\mathbf{K} \), and \( \mathbf{V} = x_i\mathbf{W}^\mathbf{V} \). \( \mathbf{W}^\mathbf{Q} \), \( \mathbf{W}^\mathbf{K} \), \( \mathbf{W}^\mathbf{V} \in \mathbb{R}^{D \times D} \) are parametric projection matrices, and \( D \) denotes the input token dimension. Finally, the output is passed through a linear layer and added to the residual connection
\begin{equation}
\text{SelfAttn}(x_i) = \text{Linear}(\text{attention}(x_i)) + x_i.
\end{equation}
To alleviate the computational cost increase brought by 3D processing, we replace vanilla global attention with window attention or plane attention. Figure \ref{fig:attention} provides a visual explanation of these operations.

For window attention, we use a window size of \(w \times w \times w\), reducing the input token sequence length \(L\) to \(w^3\)
\begin{equation}
    x_i = \text{Reshape}(B\cdot \frac{L}{w^3}, w^3, D)(x_i),
\end{equation}
where \(x_i\) is the input tokens, \(\text{Reshape}(B\cdot \frac{L}{w^3}, w^3, D)\) denotes dividing the original \(L\) input tokens into \(\frac{L}{w^3}\) groups, each of length \(w^3\), where attention operations are performed only within each group of \(w^3\) tokens. Since the time complexity of the attention operation is quadratic in the input token length, the Reshape operation reduces the time complexity from \(O(L^2)\) to \(O(Lw^3)\). In our experiments, \(w = 4\), thus reducing the time complexity by a factor of 8 (with \(L = 512\)). 

For plane attention, similar to window attention, we group tokens based on their spatial positions
\begin{equation}
    x_i = \text{Reshape}(B\cdot \frac{L}{X}, Y\cdot Z, D)(x_i),
\end{equation}
where \(\text{Reshape}(B\cdot \frac{L}{X}, Y\cdot Z, D)\) ensures that tokens only in the \(yOz\) plane can attend to each other. Note that tokens in the \(yOz\) plane may correspond to multiple flow fields because of the 3D patchify operation at the beginning of the model. In one self-attention sub-layer, we repeat plane attention three times, corresponding to the \(yOz\), \(xOz\), and \(xOy\) planes, as shown in Figure \ref{fig:attention}. These simple modifications to vanilla attention operation introduce inductive biases into the model, leveraging the locality of flow fields to enhance computational efficiency while maintaining performance.

Following the self-attention, cross-attention is applied to incorporate the 2D flow field information. Additionally, AdaLN and Scale, as shown in Figure \ref{fig:pipeline}, also serve as methods for injecting flow field information, which we will describe in detail later. The final feed-forward network (FFN) consists of two linear layers with an activation function in between, similar to \cite{vaswani2017attention}.

\textbf{Conditioning.}\quad We adopted a triple-stream conditioning mechanism. In the first stream, we pad the 2D flow field to 3D to align with the model’s input, then channel-concatenate it with the 3D flow field being denoised. This approach directly feeds the detailed information from the 2D flow field into the model, ensuring that the local details of the reconstructed flow field align with the corresponding 2D slices.

For the second conditioning stream, we use CLIP's image encoder to extract features \(F_{p_i}\) for all 2D planes \(p_i\) in the set \(P\). These features are concatenated to form the plane feature embedding \(F_P\), which is then projected to the dimension of the input tokens using an MLP. Next, the timestep embedding \(t_{emb}\) and plane position embedding \(E_P\) are added to obtain the conditional information embedding
\begin{equation}
    c(t, P, E_P) = t_{emb} + \lambda_1 F_P + \lambda_2 E_P,
\end{equation}
where \(\lambda_1\) and \(\lambda_2\) are two learnable weight parameters, both initialized to \(1\). Note that we use the global semantic token from the 2D flow field, specifically CLIP’s class token \( \mathbf{f}_{\text{cls}} \), to capture its global features. Following DiT \cite{peebles2023scalable}, we input the conditional information embedding \(c\) into the model through adaptive layer norm.

Assuming \(x_i\) is a sequence of vectors in the \(i\)-th sub-layer of the Transformer, adaLN with embedding \(c\) is defined as
\begin{equation}
    \text{adaLN}_i(x_i) = (1 + \gamma_i) \cdot \text{LN}(x_i) + \beta_i,
    \end{equation}
    \begin{equation}
    \beta_i, \gamma_i = \text{MLP}^{\text{Scale,Shift}}_i(c),
\end{equation}

where \(\gamma_i\) and \(\beta_i\) are the scale and shift parameters, and \(\text{LN}\) is the layer normalization. adaLN is applied to each  sub-layers of the Transformer layer.

Additionally, a scale \(\alpha_i\) output by another MLP is applied to \(x_i\) prior to the residual connections in each sub-layer
\begin{equation}
    \text{Scale}_i(x_i) =  \alpha_i \cdot x_i,
    \end{equation}
    \begin{equation}
    \alpha_i = \text{MLP}^{\text{Scale}}_i(c).
\end{equation}

We zero-initialize these MLPs to accelerate training \cite{goyal2017accurate}. 

Although the global token effectively represents the information of the flow field, it cannot capture the full extent of the flow. To provide more comprehensive flow information to the model, we use the CLIP image encoder's final layer output \( \mathbf{F}_{\text{last}} = \{\mathbf{f}^i\}_{i=1}^K \) as input for the third conditioning stream via cross-attention. This output has been proven to offer high-fidelity representations in numerous studies.

\section{Experiments}
\subsection{Experimental Setup}

\textbf{Datasets.}\quad In our experiments, we primarily focus on two different datasets, corresponding to the incompressible Navier-Stokes (INS) equations and the compressible Navier-Stokes (CNS) equations.

The incompressible Navier-Stokes equations are a simplified version of the original equations, assuming that the fluid density is independent of pressure. The equations are given by
\begin{equation}
    \nabla \cdot \mathbf{u} = 0,
    \rho \left(\frac{\partial \mathbf{u}}{\partial t} + \mathbf{u} \cdot \nabla \mathbf{u}\right) = -\nabla p + \mu \nabla^2 \mathbf{u} + \mathbf{f}
\end{equation}
where \(\mathbf{u}\) is the velocity, \(\rho\) is the density, \(p\) is the internal pressure, and \(\mathbf{f}\) represents external forces. We use a dataset from \cite{yousif2023deep}, consisting of a turbulent channel flow at friction Reynolds numbers \(Re_\tau = 180\).

The compressible Navier-Stokes equations are expressed as
\begin{equation}
    \frac{\partial \rho}{\partial t}  + \nabla \cdot (\rho \mathbf{u}) = 0,
\end{equation}
\begin{equation}
    \rho (\frac{\partial  \mathbf{u}}{\partial t}  + \mathbf{u} \cdot \nabla \mathbf{u}) = - \nabla p + \eta \nabla^2  \mathbf{u} + (\zeta + \eta/3) \nabla (\nabla \cdot \mathbf{u}),
\end{equation}
\begin{equation}
    {\partial   (\epsilon + \frac{\rho u^2}{2} )}/{\partial t} + \nabla \cdot \left[ \left( \epsilon + p + \frac{\rho u^2}{2} \right) \mathbf{u} - \mathbf{u} \cdot \mathbf{\sigma}' \right] = 0,
\end{equation}
where \(\epsilon= p/(\Gamma - 1)\)  is the internal energy density, \(\Gamma = 5/3\), \(\sigma'\) is the stress tensor, and \(\eta\), \(\zeta\) are the shear and bulk viscosity, respectively.
Our dataset is from \cite{takamoto2022pdebench}, with shear viscosity \(\eta\) of \(1 \times 10^{-8}\) and bulk viscosity \(\zeta\) of \(1 \times 10^{-8}\), Mach number as 1.0, and turbulence initial conditions.

Both datasets are generated using direct numerical simulation (DNS). For the INS dataset, we employ two different data partitioning methods to test the model's interpolation and extrapolation capabilities, denoted as INS(INT) and INS(EXT), respectively. Specifically, each flow field in the INS dataset has a corresponding time step. For INS(INT), we randomly split data from different time steps into training and test sets, whereas for INS(EXT), we use data from the first 80\% of time steps for training and the remaining time steps for testing. Since \(\Delta t = 0.07\), the flow fields at consecutive time steps are similar, allowing INS(INT) to evaluate the model’s ability to reconstruct 3D flow fields when ample data is available. In contrast, INS(EXT) assesses the model’s generalization capability across different time steps. For the CNS dataset, we only test the model's extrapolation capability. In addition, we also test our model's capability to reconstruct flow fields around geometric objects (Section \ref{geo}).

\textbf{Metrics.}\quad We use normalized root-mean-square error (nRMSE) measure the accuracy of the 3D flow field reconstruction
\begin{equation}
    \text{nRMSE} =  \frac{\left \| u_{pred} - u_{true} \right \|_2}{\left \| u_{true} \right \|_2} , \label{l2RE}
\end{equation}

and use the peak signal-to-noise ratio (PSNR) to assess the quality of the reconstructed 2D slices of the flow field
\begin{equation}
    \text{PSNR} = 10 \cdot \log_{10}{\frac{\text{MAX}^2(u_{true})}{\text{MSE}(u_{true}, u_{pred})}} . \label{PSNR}
\end{equation}
We also adapt the structural similarity index measure (SSIM) to better assess the detailed aspects of the reconstructed flow field
\begin{equation}
    \text{SSIM} = \frac{(2\mu_{u_{true}} \mu_{u_{pred}} + C_1)(2\sigma_{u_{true} u_{pred}} + C_2)}{(\mu_{ u_{true} }^2 + \mu_{u_{pred}}^2 + C_1)(\sigma_{u_{true}}^2 + \sigma_{ u_{pred} }^2 + C_2)} , \label{SSIM}
\end{equation}
where \(C_1\) and \(C_2\) are constants, which we set to 0.01 and 0.03, respectively. We calculate SSIM with a local window of \(11\times 11 \times 11\).

\textbf{Implementations.}\quad We implemented three different model sizes: Small, Base, and Large. Table \ref{tab:models} gives details of our models. Note that only some of the global attention in the model are replaced with window attention or plane attention. 

We reconstruct the 3D velocity field \(u, v, w\). The model uses the AdamW \cite{kingma2014adam,loshchilov2017decoupled} optimizer. The learning rate is set to \(1 \times 10^{-4}\) and we use a Cosine-Annealing-LR scheduler. The batch size is 32, and we train for 10,000 epochs.

\begin{table}[h!]
    \centering
    \caption{Details of models. (*) denotes models that use window attention and plane attention.}
    \resizebox{\textwidth}{!}{%
    \begin{tabular}{lccccc}
    \toprule
    \multirow{2}{*}{\textsc{Model} } & \multirow{2}{*}{\textsc{Layers $N$}} & \multirow{2}{*}{\textsc{Hidden size $D$}} & \multirow{2}{*}{\textsc{Heads}} & \textsc{Window Attn} & \textsc{Plane Attn}\\ 
    & & & & \textsc{Layer} & \textsc{Layer} \\
    \midrule
    \textsc{Small} & 8 & 384 & 6 & N/A & N/A \\ 
    \textsc{Small*} & 8 & 384 & 6 &(1,4)& (2,5) \\ 
    \textsc{Base} & 10 & 648 & 8 & N/A & N/A \\ 
    \textsc{Base*} & 10 & 648 & 8 &(1,4,7) & (2,5,8)\\ 
    \textsc{Large} & 12 & 768 & 12 & N/A & N/A\\ 
    \textsc{Large*} & 12 & 768 & 12&(1,4,7,10) & (2,5,8,11) \\ 
    \bottomrule
    \end{tabular}
    }
    \label{tab:models}
\end{table}

\subsection{Baselines}
We briefly introduce the baseline models compared in the experiments, with detailed descriptions provided in the Appendix.

\textbf{F-FNO.}\quad The Fourier Neural Operator (FNO) \cite{li2020fourier} is a model designed for solving partial differential equations (PDEs) by learning mappings between infinite-dimensional function spaces. It leverages Fourier transforms to operate in the frequency domain, enabling efficient handling of complex continuous dynamical systems. The Factorized FNO (F-FNO) \cite{tranfactorized} introduces improvements over the original FNO, particularly in terms of model depth, to enhance its scalability. In our experiments, we used a 24-layer F-FNO with 16 Fourier modes and 64 channels.

\textbf{U-Net-mod.}\quad We compared with U-Net-mod, an improved architecture based on the traditional U-Net, designed to accommodate more complex tasks. The modifications in U-Net-mod includes a deeper network, more sophisticated convolutional kernels, alternative upsampling strategies, and the incorporation of novel regularization methods and attention mechanisms.

\textbf{Dil-ResNet.}\quad The dilated ResNet (Dil-ResNet) is proposed by \citet{stachenfeld2021learned} as a neural PDE solver, it achieves a larger receptive field with fewer convolution operations compared to traditional ResNet. However its speed is significantly slower. Therefore, we limited our model to an 8-layer Dil-ResNet with 64 channels, where each layer consists of seven 3D convolutions with dilation rates of [1, 2, 4, 8, 4, 2, 1].

For all baseline models, we padded 2D planes to three dimensions as input, with the corresponding 3D flow as output. MSE is employed as the loss function. Additionally, we report the performance of the 2D3DGAN proposed by \citet{yousif2023deep} on the INS dataset.

\begin{figure*}[ht]
    \centering 
    \includegraphics[width=6cm]{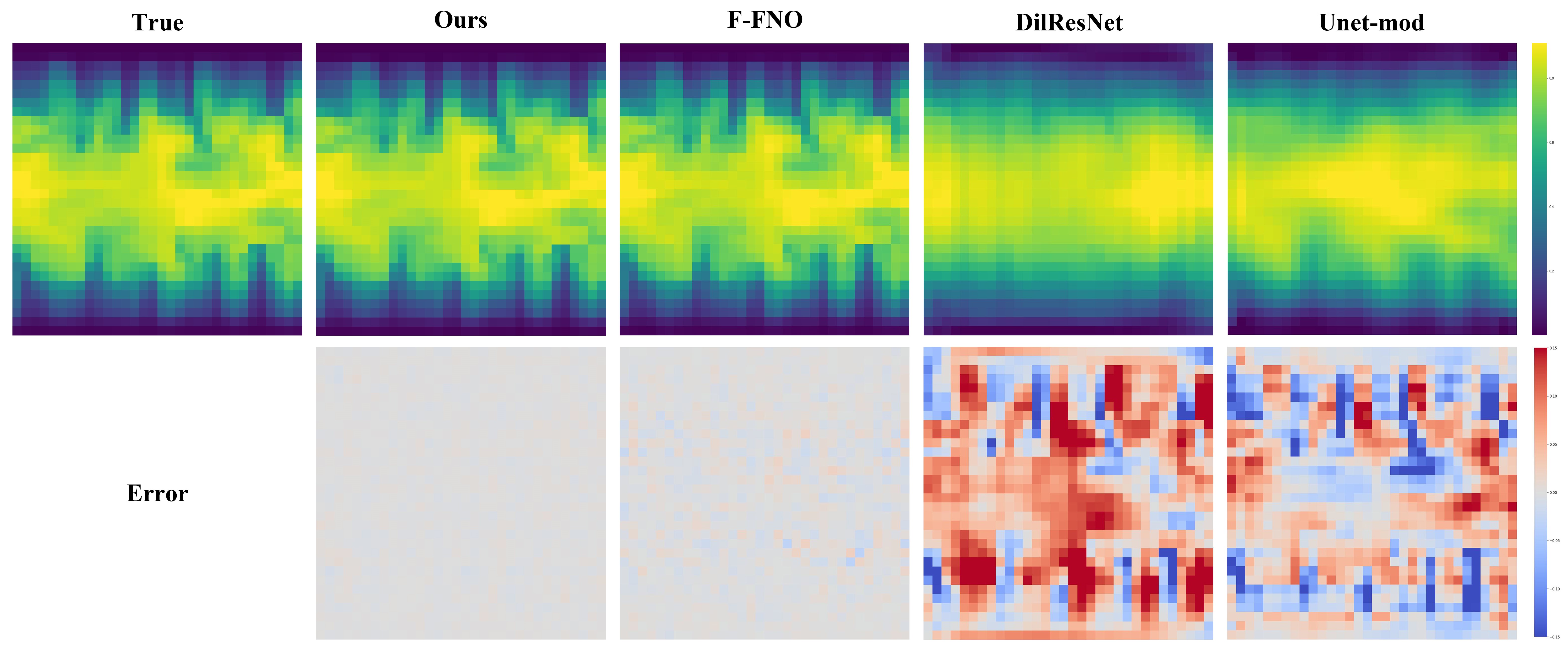}
    \includegraphics[width=6cm]{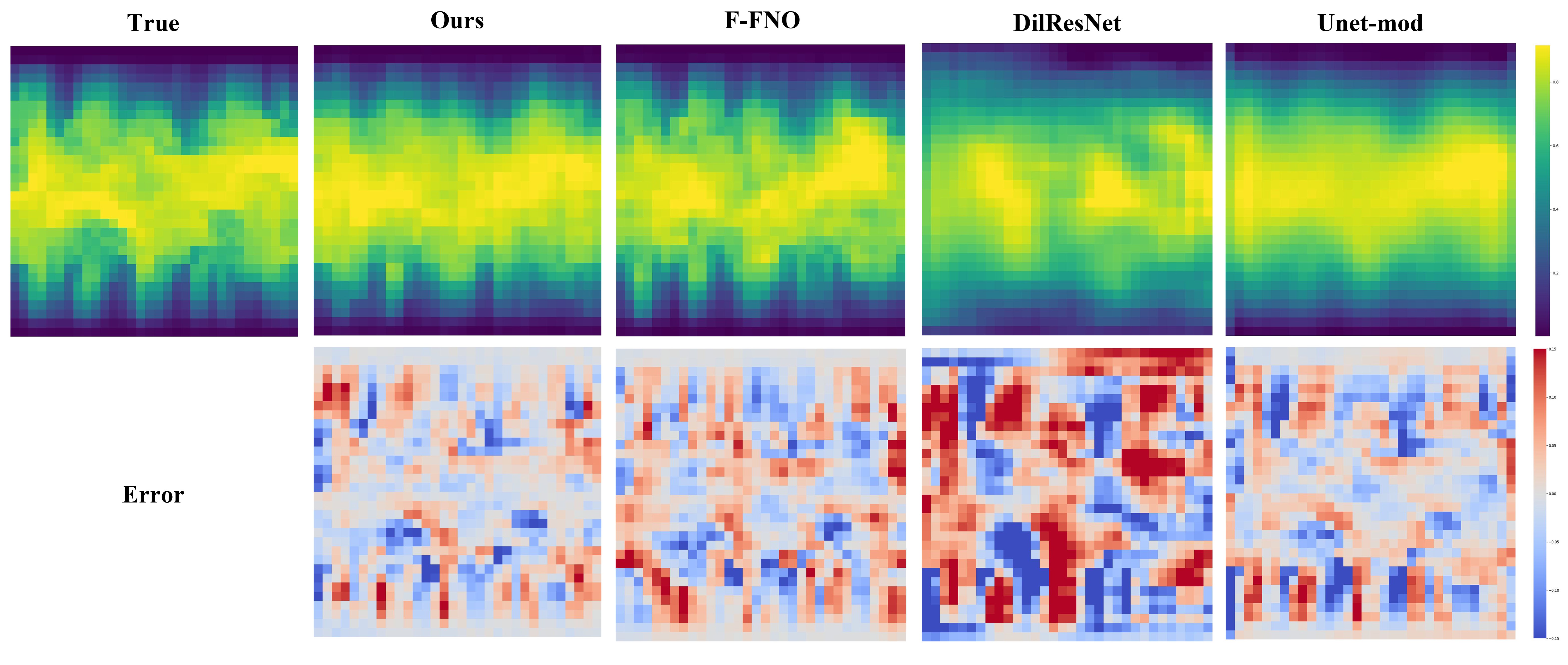}
    \includegraphics[width=6cm]{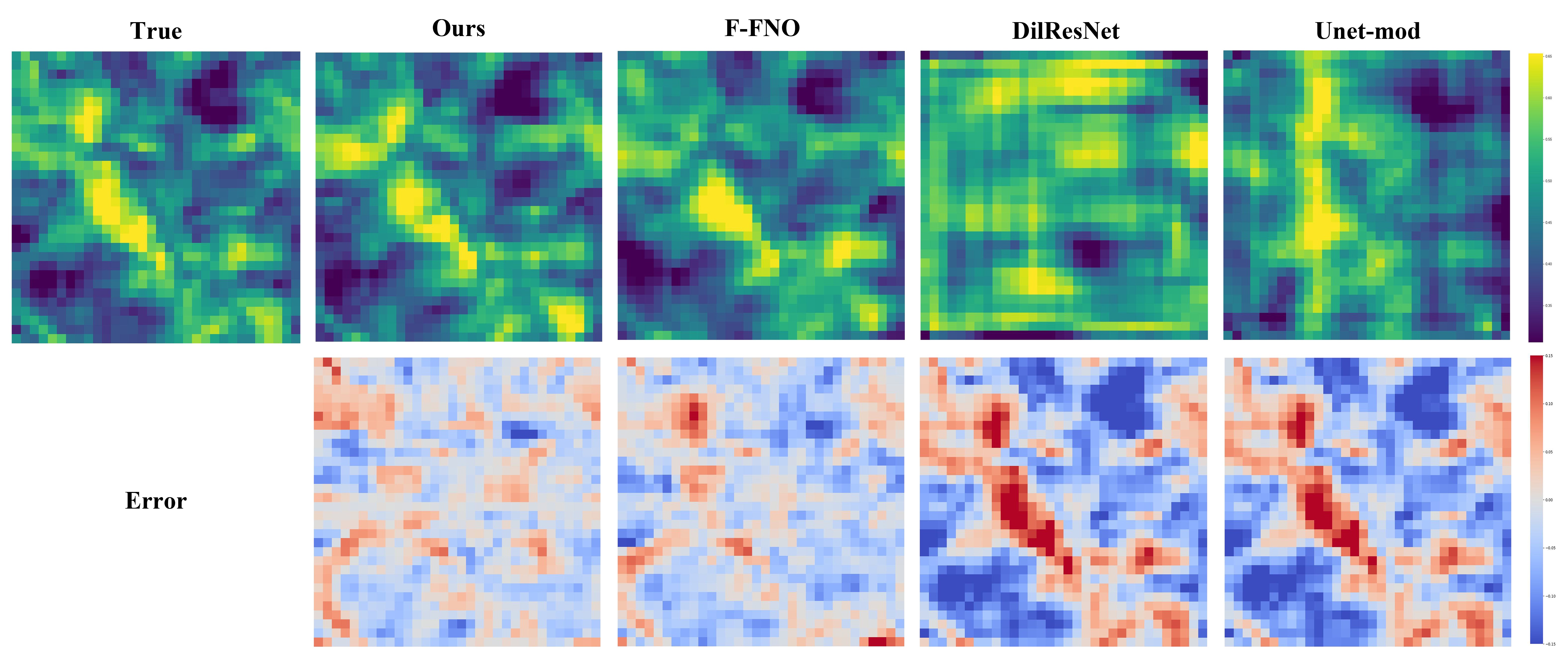}
    \includegraphics[width=6cm]{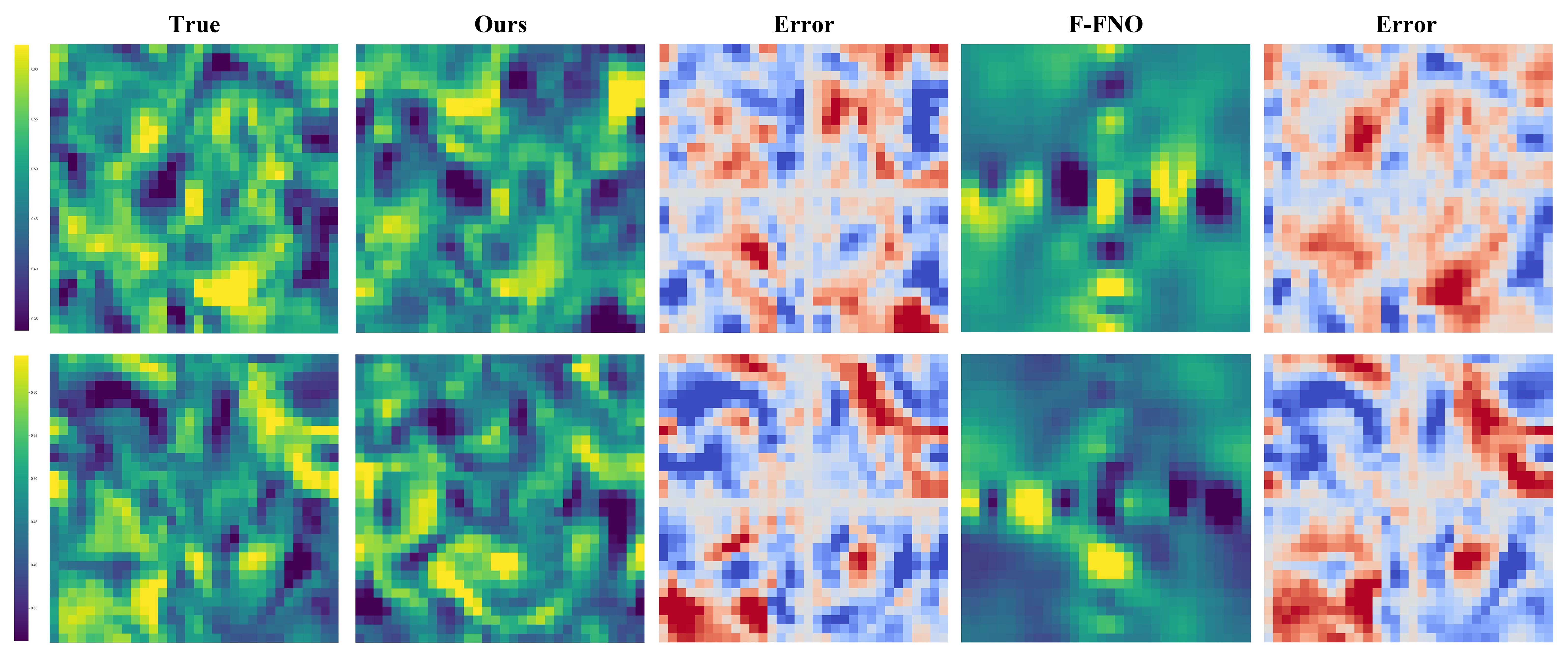}
    \caption{Visualization of results from different models on the INS(INT) (top left), INS(EXT) (top right), CNS (bottom left), and CNS \(xOy\) planes where the reference \(xOy\) plane is not provided. Note that the displayed 2D flow fields are not the inputs to the models.}
    \label{fig:result-viz}
\end{figure*}

\subsection{Results}
Table \ref{tab:results} shows the test results of the baseline models and our proposed method on three datasets. All models reconstruct the 3D flow field from two orthogonal 2D flow field planes. Figure \ref{fig:result-viz} presents visualizations of flow fields reconstructed by different models, with the flow field planes taken from planes adjacent to the reference 2D planes used for reconstruction.

Across all three metrics, our method achieves the best results on all datasets, especially on the INS(INT) dataset. This demonstrates that when the training data covers all flow field information, our method can very accurately reconstruct the 3D flow field from 2D observations. Based on the metrics, our Large model may exhibit overfitting, likely due to the lack of measures to prevent overfitting, such as dropout. In terms of flow field visualization, other methods generally fail to reconstruct the detailed information of the flow field, which is also evident from the SSIM metric. Surprisingly, the flow field reconstructed by F-FNO also appears highly realistic and is not much worse than ours.

In the other two datasets, where we test the model's extrapolation capability, nearly all models show a decline in test metrics. Although our method does not outperform others significantly in these metrics, its lead in SSIM indicates superior reconstruction of flow field details, which is corroborated by the visualized results. 

Additionally, as shown in the bottom right of Figure \ref{fig:result-viz}, when flow field information for a specific 2D plane, such as the \(xOy\) plane, is not provided, F-FNO fails to reconstruct the corresponding flow field, resulting in meaningless noise. In contrast, while our method may not achieve the same accuracy as when plane information is available, it can still reconstruct a meaningful flow field.

\begin{table}[h!]
\caption{Summary of the three test metrics on different datasets. An asterisk (*) denotes models that use window attention and plane attention. The best performance is shown in bold.}
\centering
\small
\resizebox{\textwidth}{!}{%
\begin{tabular}{@{}lccccccccccc@{}}
\toprule
\multirow{2}{*}{\textsc{Model}} &\multicolumn{3}{c}{\textbf{INS(INT)}} & \multicolumn{3}{c}{\textbf{INS(EXT)}} & \multicolumn{3}{c}{\textbf{CNS}}  \\ 
 & nRMSE(\(\downarrow\)) & PSNR(\(\uparrow\)) & SSIM(\(\uparrow\)) & nRMSE(\(\downarrow\)) & PSNR(\(\uparrow\)) & SSIM(\(\uparrow\)) & nRMSE(\(\downarrow\)) & PSNR(\(\uparrow\)) & SSIM(\(\uparrow\)) \\ \midrule
\textsc{F-FNO} & 0.0352& 39.8618 & 0.9630   & 0.1421&	22.4728 &	0.7966 & 0.1230 & 24.2100 & 0.7969 \\ 
\textsc{Unet-mod} & 0.1267& 23.4915 & 0.7861   & 0.1245&	23.6319&	0.7860 & 0.1233 &	24.8443& 0.7517 \\ 
\textsc{DilResNet} & 0.1354& 22.8427 & 0.7740   & 0.1651&	21.2144&	0.7424 & 0.1250 &	24.7098& 0.7423 \\ 
\textsc{2D3DGAN} & 0.3630 & - & -   & - &	- &	- & - &	- & - \\ \midrule
\textsc{Small (Ours)} & 0.0653& 29.1077 & 0.9567   & 0.1298&	22.9346&	0.8132 & 0.1225 &	24.4753& 0.8349 \\ 
\textsc{Base (Ours)}  &\textbf{0.0053} &\textbf{51.0158}  & \textbf{0.9997}  &  0.1331&	22.9346 & 0.8105 &0.1180	&24.8012	&\textbf{0.8410} \\ 
\textsc{Large* (Ours)} & 0.0121& 43.7522 &0.9992   & \textbf{0.1229} & \textbf{23.7372} & \textbf{0.8119}  & \textbf{0.1154} &	\textbf{25.0022}	&0.8361 \\
\bottomrule
\end{tabular}%
}
\label{tab:results}
\end{table}

\begin{figure*}[h]
    \centering 
    \includegraphics[width=12cm]{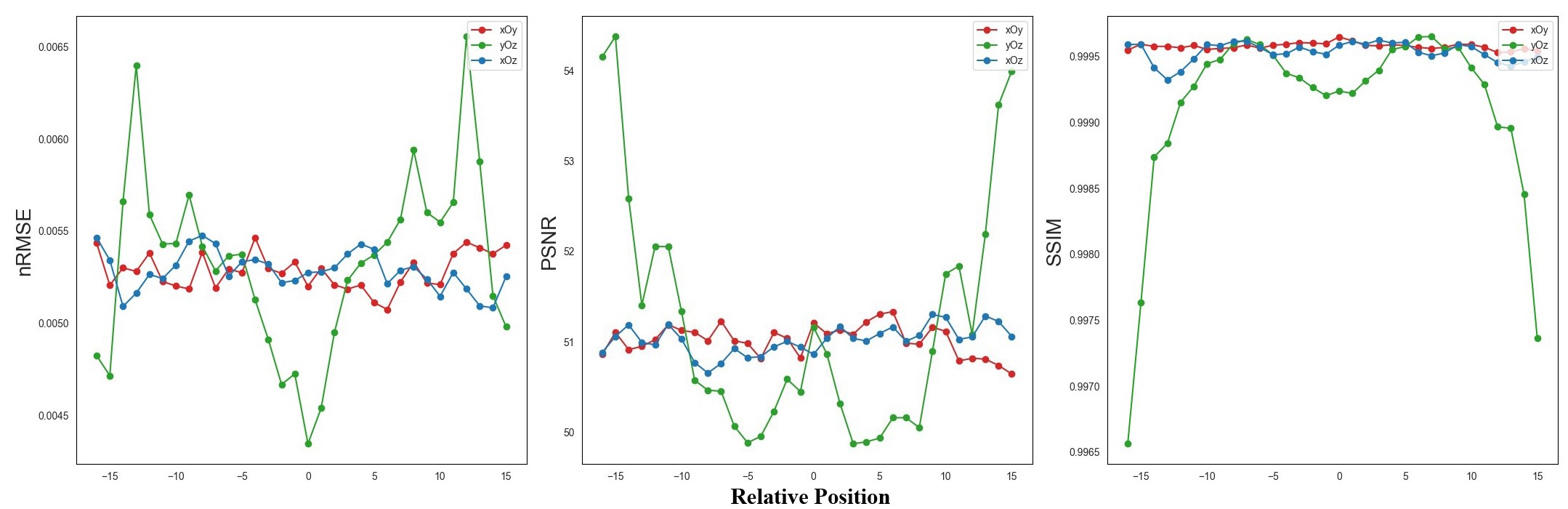}
    \caption{Relative positions to the center of the 3D flow field versus three test metrics on INS(INT).}
    \label{fig:INS(INT)-chart}
\end{figure*}

We plotted three different test metrics of the model's reconstruction results across three planes (\(xOy, yOz, xOz\)) on the INS(INT) dataset, as shown in Figure \ref{fig:INS(INT)-chart}. The relative position indicates the plane's location relative to the 2D plane used for reconstruction, i.e., the center of the XYZ cube. We can see that our model maintains high reconstruction quality across different spatial positions, although the reconstruction results on the yOz plane show slight fluctuations.

\begin{figure*}[h]
    \centering 
    \includegraphics[width=12cm]{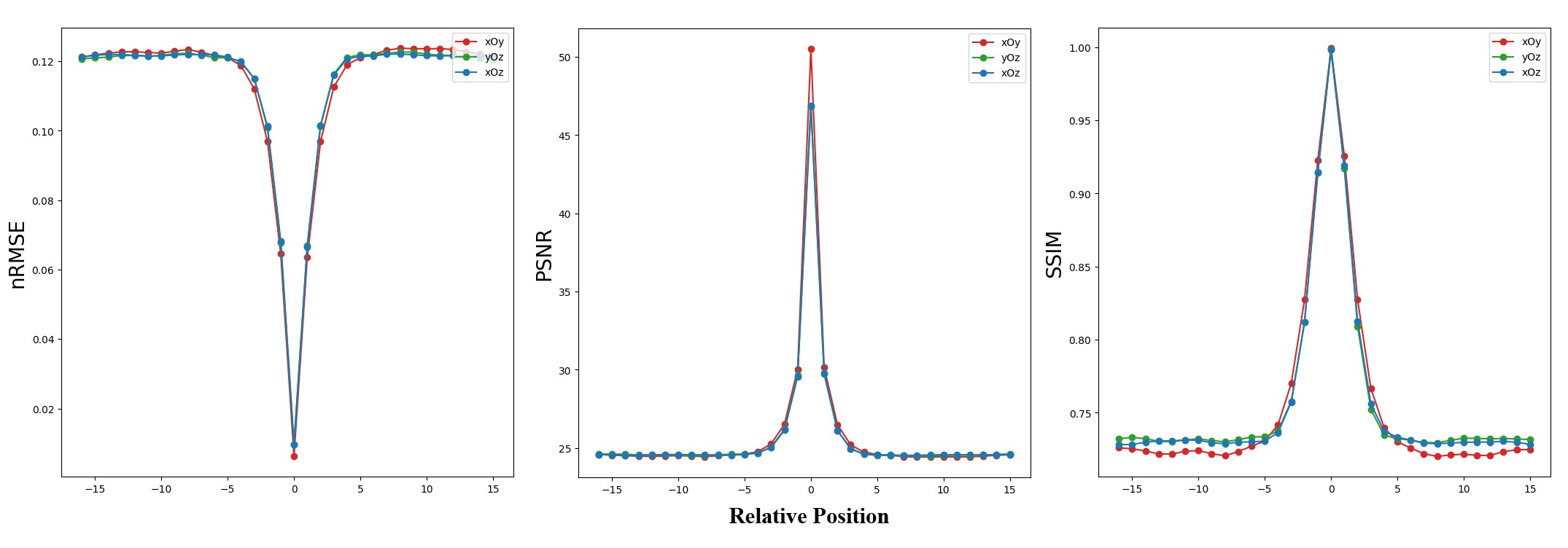}
    \caption{Relative positions to the center of the 3D flow field versus three test metrics on CNS.}
    \label{fig:rp0}
\end{figure*}

Figure \ref{fig:rp0} presents the same three test metrics across three planes on the CNS dataset. It can be seen that our model reconstructs the flow field outward from the input 2D flow fields (reference flow fields). The closer the position of the 2D slices in the reconstructed 3D flow field is to the reference flow fields, the better the reconstruction effect. This is because the closer distance ensures more accurate flow field information provided by the reference flow fields. In contrast, at positions far from the reference flow fields, the reconstructed flow field has a larger error compared to the true values. This suggests that increasing the number of reference flow fields used for reconstruction and ensuring their even spatial distribution can improve the reconstruction quality of the flow field.

\subsection{Reconstructing the 3D flow around geometries}\label{geo}

\begin{figure*}[h]
    \centering 
    \includegraphics[width=2.5cm]{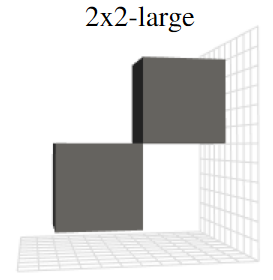}
    \includegraphics[width=8cm]{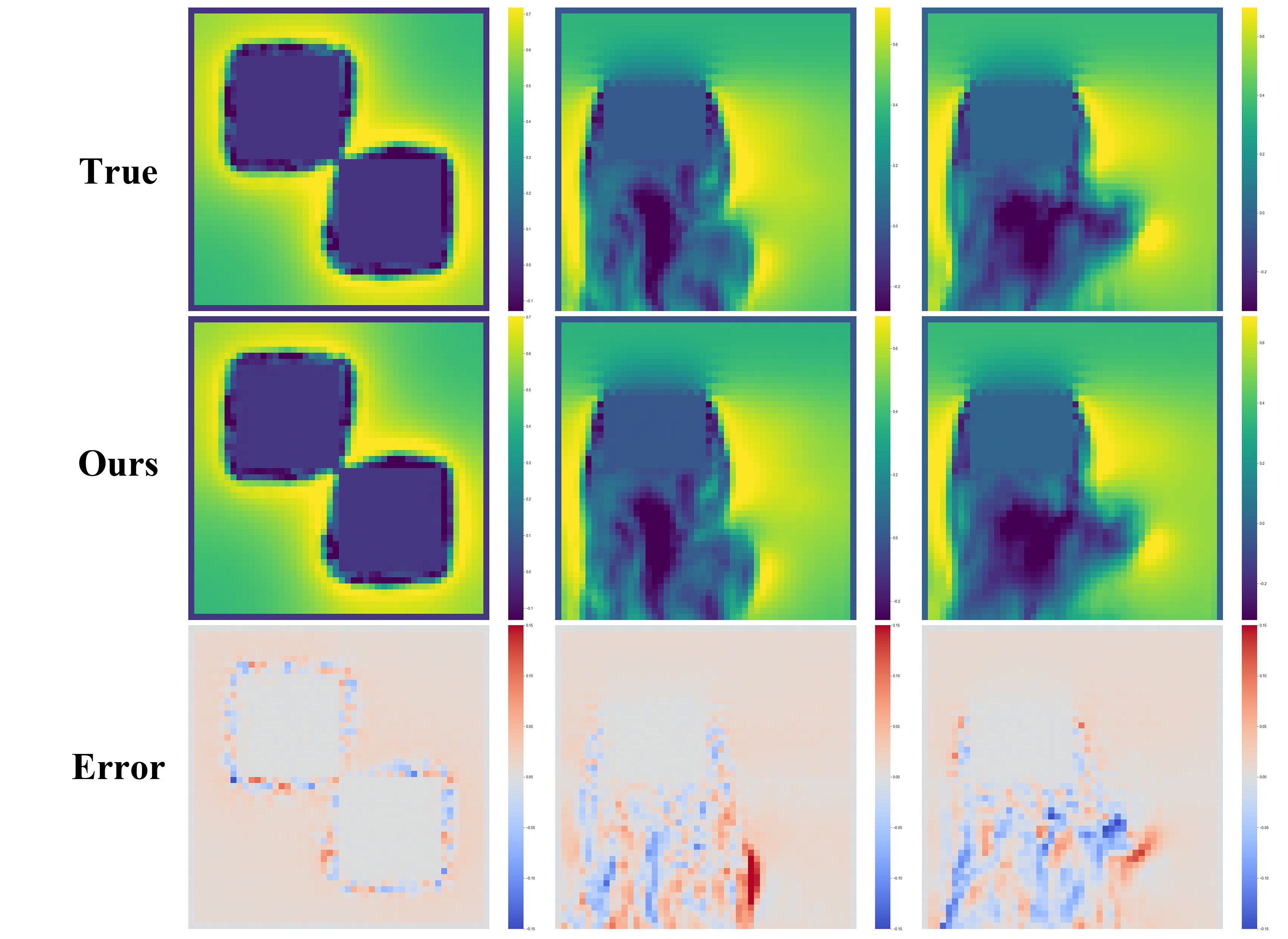}
    \caption{Visualization of reconstructing the 3D flow field around geometries. Left: 2x2 large object in dataset \cite{lienenzero}. Right: results, the velocity field \(u\).}
    \label{fig:2x2-large}
\end{figure*}

In addition to testing our model’s ability to reconstruct 3D channel turbulence, we also evaluated its performance in reconstructing flow fields around geometries. We used the 2x2-large dataset from \cite{lienenzero}, with the geometries in the dataset illustrated on the left side of Figure \ref{fig:2x2-large}. Table \ref{tab:2x2-large} presents the model's reconstruction results, while the right side of Figure \ref{fig:2x2-large} shows the corresponding visualizations. The results indicate that our model effectively handles flow fields around geometries, achieving high realism and accuracy in the reconstructions.

\begin{table}[h!]
    \caption{Corresponding metrics for the reconstruction results of the 3D flow around geometries. }
    \centering
    \small
    \begin{tabular}{@{}lccccccc@{}}
    \toprule 
    \textsc{Model} & nRMSE(\(\downarrow\)) & PSNR(\(\uparrow\)) & SSIM(\(\uparrow\)) \\ \midrule
    \textsc{Large*}  & 0.1524 & 33.4711 & 0.9793 \\
    \bottomrule
    \end{tabular}%
    \label{tab:2x2-large}
\end{table}

\subsection{Ablation Study}
In this section, we present the results of the ablation experiments.

\textbf{Number of 2D Planes.}\quad Table \ref{tab:abla-num-planes} shows the impact of the number of 2D planes used for reconstructing the 3D flow field on reconstruction quality. Adding additional 2D planes does not significantly affect reconstruction quality. The Small and Large models showed no improvement with the increase in the number of planes, while the Base model showed a slight improvement but a decrease in the SSIM. It is possible that adding just one more 2D flow field does not provide enough additional information and instead puts pressure on the model's training optimization.

\begin{table}[h]
\centering
\caption{Ablation study results on the number of planes used for reconstruction, with the best results within the same model size highlighted in bold.}
\small

\begin{tabular}{@{}lccccccc@{}}
\toprule
\multirow{2}{*}{\textsc{Model}} & \multirow{2}{*}{\textsc{\# Planes}} & \multicolumn{3}{c}{\textbf{INS(EXT)}}  \\ 
    & & nRMSE(\(\downarrow\)) & PSNR(\(\uparrow\)) & SSIM(\(\uparrow\)) \\ \midrule
\multirow{2}{*}{\textsc{Small}} & 2 & \textbf{0.1298}	&\textbf{23.1500}	&\textbf{0.8132}   \\ 
                                 & 3& 0.1343	& 22.8565	&0.7915\\ \midrule
\multirow{2}{*}{\textsc{Base}}  & 2 & 0.1331& 22.9346&	\textbf{0.8105} \\ 
                                 & 3  &  \textbf{0.1290}	&\textbf{23.2084}&	0.7975\\ \midrule
\multirow{2}{*}{\textsc{Large*}}  & 2 & \textbf{0.1229} & \textbf{23.7372} & \textbf{0.8119}\\
                                 & 3 &  0.1269&	23.3493 &0.8006\\

\bottomrule
\end{tabular}%

\label{tab:abla-num-planes}
\end{table}

\textbf{Attention.}\quad The test results for models using different attention mechanisms are shown in Table \ref{tab:abla-attention}. Replacing some of the global attention in the model with window attention and plane attention resulted in at least a 25\% improvement in training speed. For the Small and Base models, this led to a slight performance impact, while the performance of the Large model actually improved. These results validate that the attention mechanisms we adopted can significantly reduce computational costs with little to no impact on model performance.

\begin{table}[h]
\centering
\caption{Ablation study results on different attention. Better results in the same model size are shown in bold.}
\small
\resizebox{\textwidth}{!}{%
\begin{tabular}{@{}lccccccc@{}}
\toprule
 \multirow{2}{*}{\textsc{Model}} &\multirow{2}{*}{\textsc{Attention}}  & \multicolumn{3}{c}{\textbf{INS(EXT)}} & \multicolumn{1}{c}{\textsc{Training} }& \multicolumn{1}{c}{\textsc{Relative}}   \\ 
   & & nRMSE(\(\downarrow\)) & PSNR(\(\uparrow\)) & SSIM(\(\uparrow\)) & (Step/Sec)  & \textsc{Promotion (\%)}  \\ \midrule
\multirow{2}{*}{\textsc{Small}} & Global & \textbf{0.1298}	&\textbf{23.1500}	&\textbf{0.8132} & 2.97 &N/A\\ 
                                 & Win\&Plane & 0.1389	& 22.5629	&0.7887 &3.80 &27.9\\ \midrule
\multirow{2}{*}{\textsc{Base}}  & Global & \textbf{0.1331}& \textbf{22.9346}&	\textbf{0.8105} &1.50 &N/A\\ 
                                 & Win\&Plane  &  0.1342&	22.8624&	0.7930 & 1.94 &31.4 \\ \midrule
\multirow{2}{*}{\textsc{Large}}  & Global & 0.1274&	23.3126& 0.7972 &1.12 &N/A \\
                                 & Win\&Plane &  \textbf{0.1269}&	\textbf{23.3493} &\textbf{0.8006} &1.56 & 39.3 \\

\bottomrule
\end{tabular}%
}
\label{tab:abla-attention}
\end{table}

\begin{figure*}[ht]
    \centering 
    \includegraphics[width=12cm]{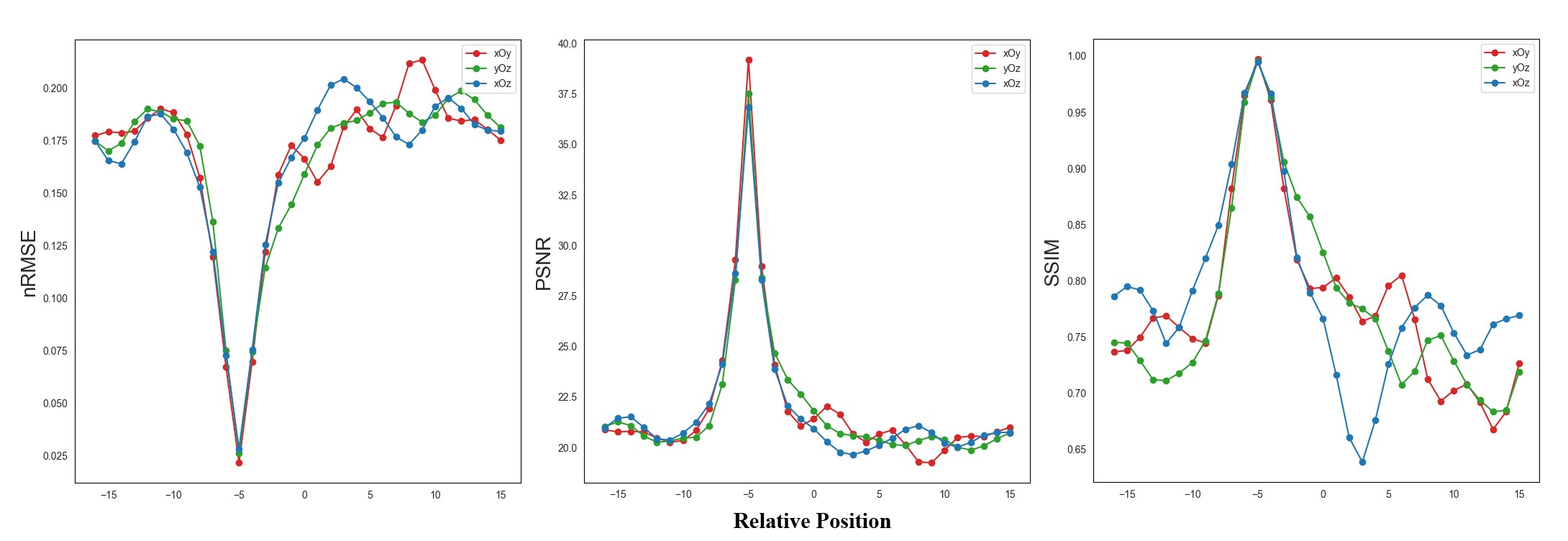}
    \caption{Relative positions to the center of the 3D flow field versus three test metrics. We use three 2D planes located at a relative position of -5 from the center of the flow field for reconstruction on CNS.}
    \label{fig:rp1}
\end{figure*}

\textbf{Plane Position Embedding.}\quad We use three 2D planes located at a relative position of -5 from the center of the flow field for reconstruction, with plane position embedding providing the positional information of the 2D planes to the model. Figure \ref{fig:rp1} shows the metrics for the reconstructed flow. It can be observed that the reconstruction quality is high for flow fields near the relative position of -5, demonstrating that our plane position embedding successfully enables the model to reconstruct 3D flow fields from 2D planes at different locations.

\textbf{Pacth Size.}\quad Table \ref{tab:abla-patch-size} presents the ablation study results on patch size. It can be observed that a smaller patch size does not significantly improve the overall quality of the flow field but is effective in reconstructing the details of the flow field. We also found that the model fails to converge during training when the patch size is too large.

\begin{table}[h!]
\caption{Ablation study results on patch size. Experiment conducted on \textsc{Large*}. The model fails to converge when patch size is 8. }
\centering
\small

\begin{tabular}{@{}lccccccc@{}}
\toprule
\textsc{Patch} & \multicolumn{3}{c}{\textbf{CNS}}  \\ 
\textsc{Size} & nRMSE(\(\downarrow\)) & PSNR(\(\uparrow\)) & SSIM(\(\uparrow\)) \\ \midrule
2 & 0.1184	&24.8302	& \textbf{0.8536}  \\ 
4  & \textbf{0.1154} &	\textbf{25.0022}	&0.8361 \\ 
8  & N/A & N/A & N/A \\

\bottomrule
\end{tabular}

\label{tab:abla-patch-size}
\end{table}

\section{Conclusion}
In this work, we propose a Diffusion Transformer-based method for reconstructing 3D flow from 2D data. Additionally, we use plane position embeddings to provide the model with the positional information of the 2D flow fields, enabling the reconstruction of 3D flow from 2D flows at arbitrary locations, thereby enhancing reconstruction flexibility. By replacing conventional global attention with window attention and plane attention, we significantly reduce the computational costs associated with the increased token length from the additional dimension, without substantially compromising model performance. Our experiments demonstrate that our model can efficiently and accurately reconstruct 3D flow fields from 2D flow data, producing realistic 3D flows. However, our approach treats flows reconstruction as a generative task, ignoring the underlying physical significance. A potential future improvement could involve embedding the corresponding physical information of the flow into the model.

\medskip

\small
\bibliographystyle{plainnat}
\bibliography{main}

\end{document}